\documentclass[12pt]{iopart}
\usepackage{iopams}
\usepackage{setstack}
\usepackage{graphicx}
\begin{document}

\title[The Early History of the Integrable Chiral Potts Model
and the Odd-Even Problem]{The Early History of the Integrable
Chiral Potts Model and the Odd-Even Problem}

\author{Jacques H H Perk}

\address{Department of Physics, Oklahoma State University, \\
145 Physical Sciences, Stillwater, OK 74078-3072, USA}
\ead{perk@okstate.edu}
\vspace{10pt}
\begin{indented}
\item[]\today
\end{indented}

\begin{abstract}
In the first part of this paper I shall discuss the round-about
way of how the integrable chiral Potts model was discovered
about 30 years ago. As there should be more higher-genus models
to be discovered, this might be of interest. In the second
part I shall discuss some quantum group aspects, especially
issues of odd versus even $N$ related to the Serre relations
conjecture in our quantum loop subalgebra paper of 5 years
ago and how we can make good use of coproducts, also borrowing
ideas of Drinfeld, Jimbo, Deguchi, Fabricius, McCoy and Nishino.
\end{abstract}

%
\vspace{2pc}

\def\vp{^{\vphantom{y}}}
\def\bx{\mathbf{x}}
\def\bX{\mathbf{X}}
\def\bZ{\mathbf{Z}}
\def\wb{{\overline W}}
\def\bbZ{\mathbb{Z}}
\font\bfra=eufb10
\def\bfrak#1{\mbox{\bfra#1}}
\font\myss=cmss12
\def\A{\mbox{\myss A}}
\def\B{\mbox{\myss B}}
\def\C{\mbox{\myss C}}
\def\D{\mbox{\myss D}}
\def\X{\mbox{\myss X}}
\def\Y{\mbox{\myss Y}}
\def\Z{\mbox{\myss Z}}
\def\xy{\textstyle{x\over y}}
\def\XY{{\displaystyle{x\over y}}}
\def\q{{q^{-1}}}

\section{Introduction}

It is a great honor to be asked to contribute to the special issue honoring
the 75th birthday of Professor Rodney Baxter and it may be proper to finally
give a review of the early history of how the integrable chiral Potts model
came into being. First of all, Baxter has given many important contributions
to this topic, as is also evidenced by his contribution in this special
issue \cite{Bax75}. Secondly, the early development of the theory went through
many unexpected twists and turns and missed opportunities not reported in the
published papers; reporting on these should encourage others seeking to discover
new solvable models and further properties of existing ones.

\section{Part 1: Early history of the integrable chiral Potts model}
\subsection{Discovery of the Yang--Baxter integrable chiral Potts model}

It really started in the summer of 1986, when Barry McCoy, Mulin Yan,
Helen Au-Yang and I decided to start a project on parafermions. The work of
Zamolodchikov and Fateev \cite{ZF} seemed to us to relate parafermionic
conformal quantum field theory to the phase transition separating regimes I
and II in the RSOS model of Andrews, Baxter and Forrester \cite{ABF},
especially, since Huse \cite{Huse} had already suggested that this transition
is in the universality class of the chiral clock model \cite{Ostlund,Huse2}.
We split the project into two parts: Au-Yang and Yan were to look at a preprint
of Alcaraz and Lima Santos, now published as \cite{ALS}, while McCoy and I
would try to find Painlev\'e-type equations that could be used to describe
parafermionic correlations, extending the conformal theory into the massive
regime, and also try a variety of perturbation expansions of correlation
functions to use for that purpose.

Au-Yang soon discovered that the full 3-state self-dual case solves the
quantum Lax pair equations, as formulated by Bashilov and Pokrovsky
\cite{BP}: There are no conditions on the corresponding spin-chain
Hamiltonian for a family of transfer matrices commuting with it to exist,
even if the Boltzmann weights in \cite{ALS} are allowed to be chiral!
Chiral comes from the Greek word $\chi\epsilon\acute\iota\rho$, hand, so
that it means handed, not reflection invariant.\footnote{Francisco Alcaraz
told us later that he had had plans to also go in that direction, but was
warned to stay away from breaking parity by one of the authorities in the
field. This, while in some sense Wu and Wang had already introduced the
one-dimensional classical chiral Potts model in a short paper on duality
transformations \cite{WW} and the chiral clock model was introduced a few
years after that \cite{Ostlund,Huse2}.} The four of us looked briefly at
Au-Yang's solution and found that the spectral variables lie on an elliptic
curve, but that the only physical two-dimensional classical spin model was
the three-state critical Potts model. McCoy and I decided for the moment
to continue with Painlev\'e and series expansions.

Au-Yang and Yan went on to analyze the non-self-dual 3-state chiral Potts model
with Boltzmann weights $W(a-b)$ and $\wb(a-b)$ for horizontal and vertical
nearest-neighbor pair interactions of spins $a$ and $b$ with values 1 to 3
(or 0 to 2) mod 3, writing
\begin{equation}
l_n={\displaystyle\sum_{j=0}^{N-1}\omega^{-jn}W(j)\over
\displaystyle\sum_{j=0}^{N-1}W(j)},
\qquad\bar l_n={\wb(n)\over\wb(0)},\qquad\omega=\mathrm{e}^{2\pi\mathrm{i}/N},
\quad N=3,
\end{equation}
as used in the quantum Lax pair approach. A logical thing to do seemed to
rewrite the star-triangle equations \cite{Baxterbook,PA} in these variables,
assuming three versions of the weights for the three different positions in
these equations, namely $W$ and $\wb$, $W'$ and $\wb{}'$, or $W''$ and
$\wb{}''$. As $l_n$ involves a discrete Fourier transform, such a transform
was also applied to the star-triangle equations, which then became, (with
proper normalizations of all $W$'s and $\wb$'s, which we can choose to be
$\sum_{j=1}^{N-1} W(j)=\overline{W}(0)=1$),
\begin{equation}
V_{ab}\overline X_b=\overline V_{ba}X_a\;,
\label{ST}\end{equation}
where
\begin{eqnarray}
&V_{ab}=\sum_{m=0}^{N-1}
\sum_{k=0}^{N-1}\omega^{am+bk+mk}\,l\vp_m\bar l'_k=
\sum_{k=0}^{N-1}\omega^{bk}\,W(a+k)\wb{}'(k),\\
&\overline V_{ab}=\sum_{m=0}^{N-1}
\sum_{k=0}^{N-1}\omega^{am+bk+mk}\,\bar l\vp_m l'_k=
\sum_{k=0}^{N-1}\omega^{ak}\,W(k)\wb{}'(b+k),
\end{eqnarray}
and\footnote{For general normalizations of the weights one has to absorb
the scalar factor $R$ in the star-triangle equation into $\overline X_a$.}
\begin{equation}
X_a=\sum_{k=0}^{N-1}\omega^{ak}\,l''_k=
\sum_{k=0}^{N-1}\omega^{ak}\,\wb{}''(k),
\qquad\overline X_a=\sum_{k=0}^{N-1}\omega^{ak}\,\bar l''_k=W''(a).
\end{equation}
Eliminating $X$ and $\overline X$ from (\ref{ST}), one gets the
consistency equations
\begin{equation}
\displaystyle{V_{ab}V_{00}\over V_{0b}V_{a0}}=
{\overline V_{ba}\overline V_{00}\over\overline V_{0a}\overline V_{b0}}\;,
\quad 1\le a,b\le N-1.
\label{condition}\end{equation}
Note that no difference-variable assumption is made on rapidities
(spectral variables). In the self-dual case $\bar l_n=l_n$,
$\overline V_{ab}=V_{ab}$, one can restrict oneself to $a<b$;
then for $N=3$ there is only one equation left and choosing
$l_1$ and $l_2$ arbitrarily, the equation then determines a
relation between $l'_1$ and $l'_2$, parametrizing a commuting
family of transfer matrices.

More generally, assuming the existence of a one-parameter family of
solutions leading to a commuting family of diagonal-to-diagonal transfer
matrices $T(u)$, we should by Baxter's well-known argument \cite{BaxterHam}
take the derivative of the logarithm of $T(u)$ at a shift point $u=0$, where
\begin{equation}
l_n=\alpha_n u+\beta_n u^2+\mathrm{O}\left(u^3\right),\quad
\bar l_n=\overline\alpha_nu+\bar\beta_n u^2+\mathrm{O}\left(u^3\right),
\label{hamlimit}\end{equation}
for $1\le n\le N-1$, $l_0=\bar l_0=1$, so that
\begin{eqnarray}
T(u)=T(0)\left[\mathbf{1}_{N^L}+u\mathcal{H}+
\mathrm{O}\left(u^2\right)\right].
\end{eqnarray}
It was easy to check that this leads to a spin-chain Hamiltonian
$\mathcal{H}$ of the form
\begin{equation}
\mathcal{H}=\sum_{j=1}^L\bigg(\sum_{n=1}^{N-1}\overline\alpha_n\bX_j^n+
\sum_{n=1}^{N-1}\alpha_n\bZ_j^n\bZ_{j+1}^{-n}\bigg)+c\,\mathbf{1}_{N^L},
\label{Ham}\end{equation}
where
\begin{equation}
\fl
\bZ_j=\mathbf{1}\otimes\mathbf{1}\otimes
\cdots\mathbf{1}\otimes\underset{j\rm th}
\bZ\otimes\mathbf{1}\cdots\otimes\mathbf{1},\quad
\bX_j=\mathbf{1}\otimes\mathbf{1}\otimes
\cdots\mathbf{1}\otimes\underset{j\rm th}
\bX\otimes\mathbf{1}\cdots\otimes\mathbf{1},
\end{equation}
with \cite{Syl}
\begin{equation}
\fl
\bX\equiv
 \pmatrix{     0&     0&       0&\ldots&           0&           1\cr
               1&     0&       0&\ldots&           0&           0\cr
               0&     1&       0&\ldots&           0&           0\cr
          \vdots&\vdots&  \vdots&\ddots&      \vdots&      \vdots\cr
               0&     0&       0&\ldots&           0&           0\cr
               0&     0&       0&\ldots&           1&           0\cr},
\qquad\bZ\equiv
 \pmatrix{     1&     0&       0&\ldots&           0&           0\cr
               0&\omega&       0&\ldots&           0&           0\cr
               0&     0&\omega^2&\ldots&           0&           0\cr
          \vdots&\vdots&  \vdots&\ddots&      \vdots&      \vdots\cr
               0&     0&       0&\ldots&\hskip-5pt
                              \omega^{N-2}\hskip-5pt&           0\cr
               0&     0&       0&\ldots&           0&\hskip-20pt
                               \omega^{N-1}\hskip-20pt&\hskip-3pt\cr},
\end{equation}
satisfying
\begin{equation}
\bZ\bX=\omega\bX\bZ,\qquad \omega=\exp({2\pi\mathrm{i}/N}),\quad
 \omega^N=1,
\end{equation}
and $c$ an irrelevant constant that equals 0 with the above
normalization $l_0=\bar l_0=1$.

For the non-self-dual $N=3$ case there is one condition
\begin{equation}
{\alpha_1^{\,3}+\alpha_2^{\,3}\over\alpha_1\alpha_2}=
{\overline\alpha_1^{\,3}+\overline\alpha_2^{\,3}\over
\overline\alpha_1\overline\alpha_2}.
\label{alpharel}\end{equation}
This was first derived from the quantum Lax pair approach and then
rederived using (\ref{condition}) and substituting (\ref{hamlimit}) and
a similar equation with $u$ replaced by $u'$ for $l'_n$ and $\bar l'_n$.

Expanding only $W'$ and $\wb'$ at the shift point $u'=0$ one has two
equivalent sets of $(N-1)^2$ equations determining the commutation of
a transfer matrix with a Hamiltonian:
\begin{eqnarray}
\alpha_m\sum_{k=0}^{N-1}{l_{k-m}\over l_k}\,\omega^{nk}=
\overline\alpha_n\sum_{k=0}^{N-1}{\bar l_{k-n}\over\bar l_k}\,\omega^{mk},
\quad (1\le m,n\le N-1),
\label{eql}\\
\alpha_m\sum_{k=0}^{N-1}{\overline S_{k+m}\over\overline S_k}\,\omega^{nk}=
\overline\alpha_n\sum_{k=0}^{N-1}{S_{k+n}\over S_k}\,\omega^{mk},
\quad (1\le m,n\le N-1),
\label{eqS}\end{eqnarray}
with
\begin{equation}
S_m=\sum_{k=0}^{N-1}\omega^{mk}\,l_k=W(m),\quad
\overline S_m=\sum_{k=0}^{N-1}\omega^{mk}\,\bar l_k
=\sum_{k=0}^{N-1}\omega^{mk}\,\wb(k),
\end{equation}
the Fourier duals of the $\bar l_n$ and $l_n$.
Without conditions on the $\alpha_n$ and $\overline\alpha_n$ neither
(\ref{eql}) nor (\ref{eqS}) allow a one-parameter family of transfer
matrices, unless $N=2$ or $N=3$ selfdual. Now (\ref{eql}) is linear and
homogeneous in the alphas, so that the coefficient determinant should vanish.
For $N=3$ this was one way to derive (\ref{alpharel}).

Au-Yang then proceeded to eliminate $\bar l_1$ and $\bar l_2$ from both
systems (\ref{eql}) and (\ref{eqS}) for the case $N=3$ by the Euclidean
algorithm, assuming the consistency relation (\ref{alpharel}).
In the meantime McCoy and I were still wrestling with our half of
the project, trying to find some nonlinear Painlev\'e-type equation
trying to generalize the conformal field theory equations of Zamolodchikov
and Fateev \cite{ZF} to the massive regime. We expanded some correlations
of the ABF model and tried to fit them to quadratic or cubic relations.
We also tried several extensions of the Sato--Miwa--Jimbo approach. We did
not get very far in spite of massive computations. Part of what we found
was written up at the end of 1986 \cite{McCP}, but the major spin-off
would come two years later.

Everything changed early October 1986, when Au-Yang came to us with the
curve,
\begin{eqnarray}
{\overline\alpha_1^{\,3}-\overline\alpha_2^{\,3}\over\alpha_1^{\,3}-\alpha_2^{\,3}}\,
{\alpha_1\alpha_2\over\overline\alpha_1\overline\alpha_2}\,3\sqrt{3}\,\mathrm{i}
(l_1^{\,3}+l_2^{\,3}-l_1l_2-l_1^{\,2}l_2^{\,2})l_1l_2(1-l_1l_2)\nonumber\\
=(1+l_1^{\,3}+l_2^{\,3}-3l_1l_2)
(l_1^{\,3}+l_2^{\,3}+l_1^{\,3}l_2^{\,3}-3l_1^{\,2}l_2^{\,2})\nonumber\\
+{\alpha_1^{\,3}+\alpha_2^{\,3}\over\alpha_1^{\,3}-\alpha_2^{\,3}}
(l_1^{\,3}-l_2^{\,3})(1+l_1^{\,3}l_2^{\,3}-l_1^{\,3}-l_2^{\,3}).
\label{genus10}\end{eqnarray}
None of us could recognize this curve and McCoy showed it to Sah and Kuga
in the Stony Brook mathematics department. Several days later they told
us that the genus of the curve (\ref{genus10}) was 10. This violated the
folklore that solutions of the quantum Yang--Baxter equations are
parametrized by curves of genus at most 1!

A period of extensive checking by all four of us followed and we studied the
conditions on the alphas for $N=4$, concluding that when $N$ is not prime
the solution is not unique. We found that the double Fourier transforms
of (\ref{eql}) and (\ref{eqS}),
\begin{equation}
\sum_{m=1}^{N-1}\alpha_m{l_{q-m}\over l_q}\,\omega^{-mp}=
\sum_{n=1}^{N-1}\overline\alpha_n{\bar l_{p-n}\over\bar l_p}\,\omega^{-nq},
\quad (0\le p,q\le N-1),
\label{eql2}\end{equation}
($\alpha_0=\overline\alpha_0=0$), are easier to deal with for larger $N$. One
can even subtract from  (\ref{eql2}) the same equation with $p=q=0$. Expanding
the $l_n$ and $\bar l_n$, one finds equations with only $\alpha_n$ and
$\overline\alpha_n$. Once we got a result for $N=5$, Au-Yang and I guessed a
general solution
\begin{equation}
\alpha_n={\mathrm{e}^{\mathrm{i}(2n-N)\phi/N}\over\sin(\pi n/N)},\qquad
\overline\alpha_n=
\lambda{\mathrm{e}^{\mathrm{i}(2n-N)\bar\phi/N}\over\sin(\pi n/N)},\qquad
\cos\phi=\lambda\cos\bar\phi,
\label{alphas}\end{equation}
and verified that it satisfies all equations for all $N\ge2$.
McCoy and I also did an extensive literature search and found among
others a paper by von Gehlen and Rittenberg \cite{vGR}, who had found the
special case of (\ref{alphas}) with $\phi=\bar\phi=\frac12\pi$, causing
us to present (\ref{alphas}) in the above form. We were surprised not to
have found out earlier that \cite{vGR} is cited in \cite{ALS}.

At some point early 1987 McCoy also got his student Shuang Tang involved
in the checking before the paper was submitted, as he wanted to be absolutely sure
about the conclusion before the submission of the letter \cite{AMPTY}.
Tang would be very involved in the next stage of the project. As the principal
author of the work, Au-Yang was supposed to speak about it at the Rutgers
meeting May 7-8, but she wanted me to do it \cite{Rutgers}. There were
two back-to-back talks scheduled, one by McGuire,
`There are no higher-genus solutions of the star triangle relations' and mine
`Commuting transfer matrices in the chiral Potts models and solutions of
star-triangle equations with genus larger than one.' Before the talks McGuire
and I compared notes and found no contradiction, as he had assumed that the weights
depend only on rapidity differences, forcing the genus of the rapidity manifold
to be 0 or 1. He changed his title and his talk became a good introduction
to my talk \cite[p.~407]{Rutgers}.

At Summer Research Institute Theta Functions---Bowdoin, July 1987, I reported
our results in more detail, adding several other observations that I had made,
such as the equivalence of the quantum Lax pair and star-triangle equation
approaches and that the Dolan--Grady criterion \cite{DG} implies the existence
of an Onsager algebra, and I submitted a handwritten manuscript for publication.
The proceedings came out only two years later \cite{ThFB}, so that I had to
include an update, modifying the last sentence and adding two further paragraphs,
as a lot had happened since.

\subsection{Parametrizing the $N$-state self-dual case}

At Theta Functions---Bowdoin, Barry McCoy reported on our next nearly finished
work \cite{MPTS}. We had noted that in (\ref{condition}), (\ref{eql}),
(\ref{eqS}) and (\ref{eql2}) we have $(N-1)^2$ equations in the general case,
(i.e.\ 4 for $N=3$, 9 for $N=4$), but only $\frac12(N-1)(N-2)$ equations in
the self-dual case, (or 1 for $N=3$, 3 for $N=4$, 6 for $N=5$). Therefore, we
decided first to extract the curve for the integrable manifold of the $N=4$
self-dual case using the Euclidean algorithm both by hand and by computer
using Wolfram's SMP. Tang, McCoy and I thus obtained a curve for the 4-state
self-dual case and Han Sah recognized it as a Fermat curve, mapping it to
$x^4+y^4=z^4$ in homogeneous coordinates \cite{MPTS}.

Next we applied the same method to the 5-state self-dual case using SMP,
obtaining some curves that took several computer screens to display.
But we managed to extract their rather simple common factor and using
a map patterned after the $N=4$ case, we found a parametrization of the
Boltzmann weights in terms of the curve $x^5+y^5=z^5$. It did then not take
much imagination to conjecture the answer for general $N$ in terms of
the Fermat curve $x^N+y^N=z^N$. More precisely, we wrote the product form
\begin{equation}
\frac{l_n}{l_0}=b^{2n}\prod_{k=1}^n
\frac{\omega^{-(k-1)/2}y-\omega^{(k-1)/2}z}{\omega^{-(N-k)/2}x-\omega^{(N-k)/2}z},
\quad \bar l_n=l_n,
\label{selfdual}\end{equation}
with the Fermat curve given as
\begin{equation}
b^{-N}(x^N-z^N)=b^N(y^N-z^N),\quad
b\equiv\mathrm{e}^{\mathrm{i}\phi/N},\quad \lambda=1,
\label{fermat}\end{equation}
involving a rescaling of the homogeneous coordinates $(x,y,z)$. Generically,
the genus of the curve (\ref{fermat}) is $\frac12(N-1)(N-2)$.
For $b=1$, (\ref{selfdual}) reduces to the genus-zero solution of Fateev
and Zamolodchikov \cite[equation~(11)]{FZ}. The new self-dual result was
submitted in October 1987 as part of our contribution to the Sato
Festschrift \cite{AMPT}.

It was clear that we did not have the computer power to do the next
step, the $N=4$ non-self-dual case, the same way. Nevertheless, progress
came soon after, during five weeks following a one week conference
in Canberra in November 1987, when Rodney Baxter, Helen Au-Yang and I got
together to work out the general case.

\subsection{Star-triangle equation and full parametrization}

\begin{figure}
\begin{center}
\includegraphics[width=4in]{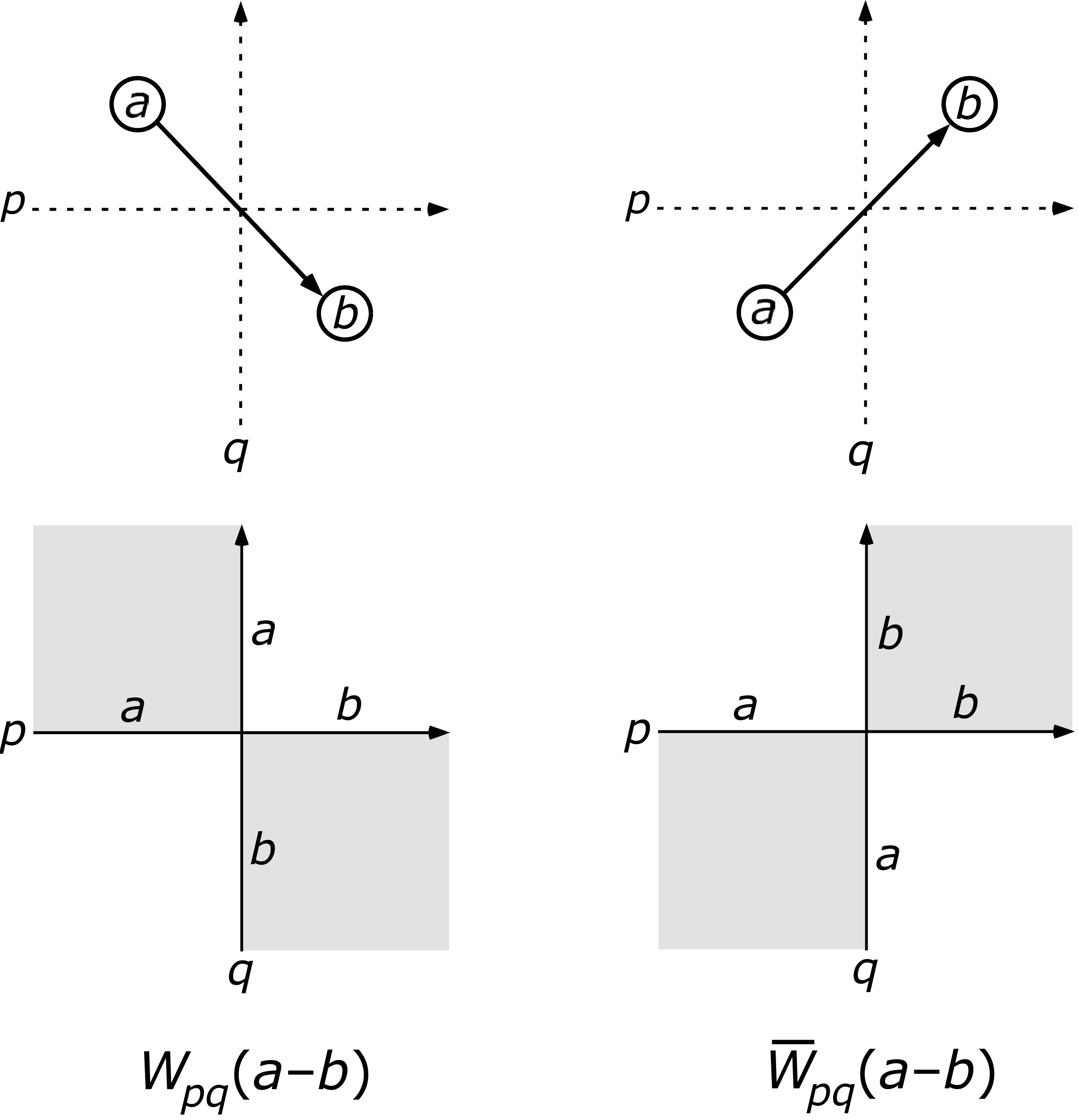}
\end{center}
\caption{The two kinds of chiral Potts model Boltzmann weights
$W_{pq}(a-b)$ and $\overline W_{pq}(a-b)$. The handedness is indicated by
arrows between the spins of values $a$ and $b$. Also given are
the oriented lines on which rapidities $p$ and $q$ live.
Row two gives the equivalent checkerboard vertex model representation.}
\label{fig1}\end{figure}

\begin{figure}
\begin{center}
\includegraphics[width=4in]{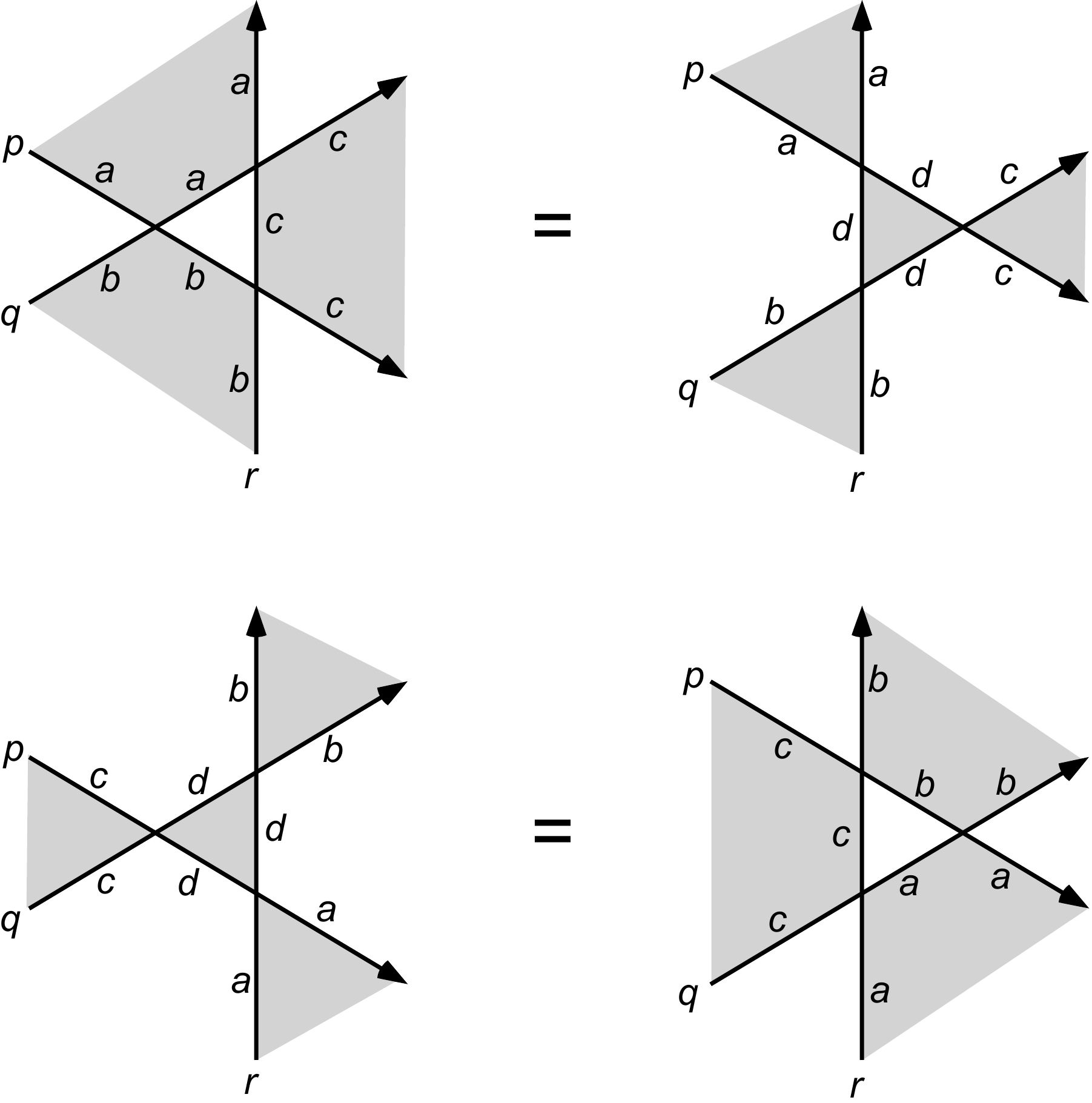}
\end{center}
\caption{The star-triangle equations represented as checkerboard Yang--Baxter
equations. For the chiral Potts model both equations (interchanging grey
and white colorings of the faces) are equivalent.}
\label{fig2}\end{figure}

The first thing to be decided was a proper notation for the Boltzmann
weights. We clearly needed to incorporate Baxter's $Z$-invariance
\cite{Baxter1978,Baxter1986}, but we decided to use notations from
\cite{AP1987}, see figure \ref{fig1}. In analogy with relativistic
scattering theory we decided to call the spectral variables $p,q,r$
rapidities and to put arrows on the rapidity lines. The chirality of
the interactions between spins, $a,b,\cdots,$ is indicated by arrows.
The Potts nature means that the weights depend on differences $a-b$
mod $N$. Not to give the same figures every time, I have here also
given the equivalent checkerboard vertex model representation,
representing the star-triangle equations as in figure \ref{fig2}.

For several days the three of us made attempts to generalize the
conjectured product form (\ref{fermat}) for the self-dual case. At some
point I proposed to look at the Ising model and to write the
theta functions of the differences of $p$ and $q$ as sums of products
of functions of $p$ and $q$ separately. We could use either
Onsager's paper \cite{Onsager} or our two papers \cite{Baxter1986,AP1987}.
Baxter and Au-Yang thought that I was crazy to guess a product from
a single factor and I was to finish the calculation I was doing. The
next day we went for an outing in Tidbinbilla, but the morning after
Baxter came up with the desired product forms \cite{BPA},
\begin{equation}
{W_{pq}(n)\over W_{pq}(0)}=
\prod_{j=1}^n\,{d_pb_q-a_pc_q\omega^j\over b_pd_q-c_pa_q\omega^j},\qquad
{\wb_{pq}(n)\over\wb_{pq}(0)}=
\prod_{j=1}^n\,{\omega a_pd_q-d_pa_q\omega^j\over c_pb_q-b_pc_q\omega^j}.
\label{weights}\end{equation}
Periodicity mod $N$, $W_{pq}(n+N)=W_{pq}(n)$, $\wb_{pq}(n+N)=\wb_{pq}(n)$,
led to the chiral Potts curve condition for the rapidities
$x_p=(a_p,b_p,c_p,d_p)$,
\begin{equation}
a_p^{\,N}+k'b_p^{\,N}=k\,d_p^{\,N},\quad
k'a_p^{\,N}+b_p^{\,N}=k\,c_p^{\,N},\quad k^2+k'{}^2=1,
\end{equation}
using a suitable normalization. We did extensive checking using Fortran on
the new little Macintosh computers at Australian National University and
the scalar factor $R$ in the star-triangle equation was guessed also from
Ising and checked by Fortran. Our letter \cite{BPA} contains many more
results, and we obtained a proof that the star-triangle equations are
satisfied using recurrences based on variations of (\ref{ST}). This proof
was published a few years later in the appendix of \cite{AP}.

\subsection{Free energy, order parameters and functional equation}

During these six weeks in Canberra, I made a brief visit to Melbourne
which was also very significant, because Paul Pearce gave me a copy of
a preprint by Bazhanov and Reshetikhin. This preprint contained a cubic
functional relation, that was omitted in the published version \cite{BR}.
As the Onsager algebra \cite{ThFB} and the paper of von Gehlen and
Rittenberg \cite{vGR} indicated that the chiral Potts model is a cyclic
version of quantum sl(2), we conjectured that chiral Potts would have a
similar functional relation. Barry McCoy thus made his students Tang and
Albertini solve the eigenvalue problem for small systems so that soon
numerical support for our conjecture would be available.

In the meantime, Baxter obtained the first free-energy result by the
``399th" method \cite{BE} to solve the Ising model and he found the
specific heat exponent to be $\alpha=1-2/N$ for the two-dimensional
classical case \cite{Bax}.

The small chain results for the von Gehlen--Rittenberg special case were
particularly simple and McCoy coined the term `superintegrable'
for this case with extra Onsager-algebra integrability, as several concepts
were being `supered' by our high-energy colleagues in Stony Brook.
Therefore, as I had some traveling to do, I wrote a couple of
self-submitting batch jobs in SMP on the three Ridge Unix computers
of the Institute for Theoretical Physics that gave us
iteratively the ground state energy of the superintegrable chain in the
commensurate phase for chains of considerable length, leading to long
series expansions for $N=3$, 4, 5 and general $N$. At some higher
orders I had to work around multiplication errors in the SMP program.

Also our earlier labor doing series expansions trying to get
Painlev\'e-type results paid off: One day McCoy told us to look again
at the paper by Howes, Kadanoff and den Nijs \cite{HKdN}, as they had
conjectured a particularly simple result for the order parameter of the
3-state superintegrable quantum chain in the ordered ground state,
namely $(1-\lambda^2)^{1/9}$ \cite[eq.~(3.13)]{HKdN}. They stated that
they had done a series expansion that reproduced this conjecture to
thirteenth order! Soon we obtained the leading terms in the expansion
of the order parameters for general $N$, so that we could generalize
their conjecture as
\begin{equation}
\langle\bZ_0^{\,k}\rangle=(1-\lambda^2)^{k(N-k)/2N^2}.
\label{order}\end{equation}
Invoking Baxter's $Z$-invariance \cite{Baxter1978}, this result should
apply also to the ordered phase of the full integrable chiral Potts model,
both for the one-dimensional quantum chain and the two-dimensional
classical case.

However, this was based on only very few terms. Luckily, just walking by,
I noted a preprint of Henkel and Lacki \cite{HL} on the very top of a pile
of discarded preprints in one of the garbage bins of the ITP. This gave
further confirmation, as Henkel and Lacki had expanded the sum of the
order parameters to one more order than we had done. Unfortunately for them,%
\footnote{Later Henkel and Lacki published \cite{HL2}, citing the preprint
of \cite{AlMPT}.}
for $N\ge4$ this sum does not have the binomial
form that it has for $N=2$ and 3. Our conjecture was submitted May 1988
in a paper with several other results \cite{AlMPT}.

Early October 1988 we received a preprint of Baxter \cite{Bax2}
in which he solved several properties of the superintegrable case by the
inversion relation. About a week later we submitted our paper \cite{AMP} with
our cubic functional equation that had been verified by Albertini for several
chain lengths using Fortran. As a result, we found that there had to exist a
commensurate-incommensurate phase transition in the superintegrable
3-state chain \cite{AMP} for $\lambda<1$, so that the conjectured phase
diagram in \cite[figure~2]{HKdN} with a Lifshitz point at $\lambda=1$ is
not quite correct.

At the Taniguchi conference, October 1988, Miwa made me present a detailed
proof that the weights (\ref{weights}) satisfy the star-triangle equation.
Each time I had filled his high-tech whiteboard he printed a copy of what
I had written. A more elaborate paper appears in the proceedings \cite{AP},
with the proof given in the appendix. McCoy presented details \cite{AMP2}
about the appearance of the incommensurate phase in the 3-state superintegrable
quantum chain. Multi-particle excitations were also studied
to estimate size of the incommensurate phase \cite{AMP3}.

\subsection{Representation theoretical understanding, outlook
and some prehistory}

There were many further developments, especially the representation theory
explanation of Bazhanov and Stroganov, valid for odd $N$ \cite{BS0,BS},
and an alternative approach valid for all $N$ \cite{BBP}. The difference
of these two approaches is discussed in the next section. These two works
made the relation with cyclic representations of quantum groups as described
later by De Concini and Kac \cite{dCK} explicit, confirming our earlier
thoughts on it.

There should be some models that have rapidities on higher-genus curves
other than the one of chiral Potts. Martins \cite{Martins} claims to have
found a model parametrized by a K3 surface recently and it would be
interesting to investigate this further.

To conclude this section, we should mention some early works of Krichever and
Korepanov, of which we were not aware until some time in 1993 when we received
some copies in Russian in the mail. In \cite{Kri81,Kri82} Krichever proved his
Theorem 1 stating that generically the genus of the curve coming from
vacuum vectors of an $N$-state model related to the six-vertex model had to
be $(N-1)^2$, but he only worked out the case $N=2$ in detail. Korepanov followed
this up studying the cases $N=3$ and greater \cite{Kor86}, discovering thus
the Boltzmann weight of some $\tau_2$ model, in agreement with Krichever's
theorem and with \cite[equation~(13)]{Kri81}, \cite[equation~(10)]{Kri82}.
This way, though unknown outside the Soviet Union for many years, Korepanov
gave the first explicit demonstration of a solution of the quantum Yang--Baxter
equation with a higher-genus parametrization, several months before the
discovery of the integrable chiral Potts model.

However, Korepanov did not discover the integrable chiral Potts model, nor
did he construct the $R$-matrix intertwining two cyclic representations.
That construction had to wait until \cite{BS0,BS}. Only with the complete
construction does one know that both the horizontal and vertical transfer
matrices of the $\tau_2$ model form commuting families with one set of
spectral parameters (rapidities) taken from the high-genus curve and the
other set from the genus-zero curve of the six-vertex model. Until
\cite{BS0,BS} the meaning of Korepanov's discovery was veiled.

Finally, in the introduction of our recent paper \cite{AP9} one can find
some other references related to parafermions that are of historical
interest, including papers on generalized Clifford algebras.

\section{Part 2: Odd or Even}

\subsection{Ising case $N=2$: Onsager algebra and Jordan--Wigner
transformation}

When $N=2$, the integrable chiral Potts model becomes the Ising model.
The chiral Potts spin-chain Hamiltonian (\ref{Ham}) reduces to
\begin{equation}
\mathcal{H}=\sum_{j=1}^L\big(\alpha_1\bZ_j\bZ_{j+1}+\bar\alpha_1\bX_j\big)
=-\sum_{j=1}^L\big(J\sigma^z_j\sigma^z_{j+1}+B\sigma^x_j\big),
\end{equation}
identifying $\bX=\sigma^x$ and $\bZ=\sigma^z$ as Pauli matrices,
and $\alpha_1=-J$, $\bar\alpha_1=-B$. This is the transverse-field
Ising chain Hamiltonian, now so popular in quantum information circles.
As said before, the connection with the Ising model for $N=2$ has
been very important for us to find the high-genus solutions of the
star-triangle equations \cite{BPA,AP}.

In the Ising case of spin-${1\over2}$, the spin operators have mixed
commutation relations: commuting if at different sites, but anticommuting
at the same site. This was first addressed in the Ising model context
by Bruria Kaufman \cite{Kaufman}, who introduced the Clifford algebra
spinors,
\begin{equation}
\fl
\Gamma_{2j-1}=\bX_1\bX_2\cdots\bX_{j-1}\bZ_j,
\quad\Gamma_{2j}=\mathrm{i}\bX_1\bX_2\cdots\bX_j\bZ_j,\quad
\Gamma_m\Gamma_n+\Gamma_n\Gamma_m=2\delta_{mn}.
\end{equation}
A generalization to spin-${1\over2}$ XXZ models at roots of unity was
introduced by Deguchi, Fabricius and McCoy \cite{DFM}. Nishino and
Deguchi \cite{ND1} found a further generalization applicable to the
superintegrable $\tau_2$-model when $N$ is odd, as required in \cite{BS}.
This was followed by a series of papers on the superintegrable $\tau_2$
and chiral Potts models \cite{AP2,ND2,AP3,AP4,AP5,AP6,AP7,AP8} for general
$N$ using \cite{BBP}. In these papers we constructed the eigenvectors in
the ground state sector and the order parameters using a generalized
Jordan--Wigner transform.

In these superintegrable models there is additional sl(2) loop group
symmetry, supporting representations of the Onsager algebra
\cite{ThFB,Onsager},
\begin{equation}
\fl
[A_j,A_k]=4G_{j-k},\qquad[G_m,A_l]=2A_{l+m}-2A_{l-m},
\qquad[G_j,G_k]=0,
\end{equation}
but with a more complicated closure relation than in the Ising model.
It should be noted that von Gehlen and Rittenberg \cite{vGR}
had already constructed the superintegrable chiral Potts quantum chain
in 1985, using the Dolan--Grady criterion \cite{DG,ThFB,Davies}, 
\begin{eqnarray}
\mathcal{H}=A_0+\lambda A_1,\quad &[A_0,[A_0,[A_0,A_1]]]=16[A_0,A_1],
\nonumber\\
&[A_1,[A_1,[A_1,A_0]]]=16[A_1,A_0].
\end{eqnarray}
which is a kind of Serre relation implying the existence of the
Onsager algebra.

\subsection{Bazhanov--Stroganov construction}
\begin{figure}
\begin{center}
\includegraphics[ width=4in]{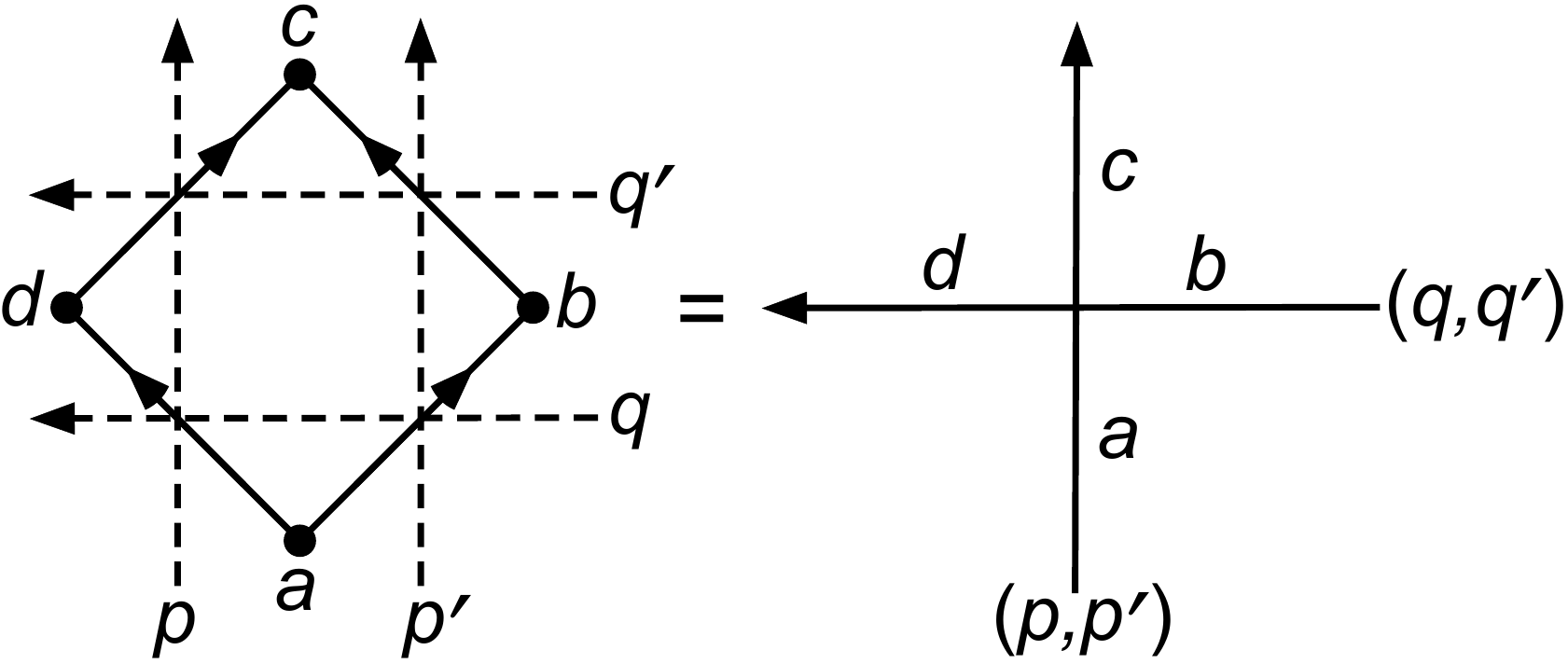}
\end{center}
\caption{Square of four chiral Potts weights (\ref{weights}) represented
by four oriented solid lines connecting the Potts spins $a$, $b$, $c$
and $d$. The four rapidity lines are drawn dashed. This is equivalent
to a vertex model with the spins living on line pieces.}
\label{fig3}\end{figure}
A quantum group construction of the integrable chiral Potts model
has first been given by Bazhanov and Stroganov \cite{BS} for odd $N$,
starting from an R-matrix of the six-vertex model, the intertwiner of
two highest-weight spin-$1\over2$ representations. They constructed
next the intertwiner of a spin-$1\over2$ and a cyclic representation
($\tau_2$-model weights). Finally, the square of four chiral Potts
weights as given in \cite{BPA} and figure~\ref{fig3} was shown to
intertwine two cyclic representations.

The more general the six-vertex R-matrix is chosen, the more easy
it is to arrive at the chiral Potts model. Korepanov \cite{Kor86}
had chosen $R^{00}_{00}=R^{11}_{11}$, $R^{01}_{01}=R^{10}_{10}$ and
$R^{10}_{01}=R^{01}_{10}$. Bazhanov and Stroganov \cite{BS} made
a special gauge choice with $R^{10}_{01}\ne R^{01}_{10}$, whereas
\cite{BBP} also ended up with $R^{01}_{01}\ne R^{10}_{10}$, see
\cite[equation~(1.7)]{AP10} for the comparison.

\subsection{Baxter--Bazhanov--Perk construction}

\begin{figure}
\begin{center}
\includegraphics[ width=4in]{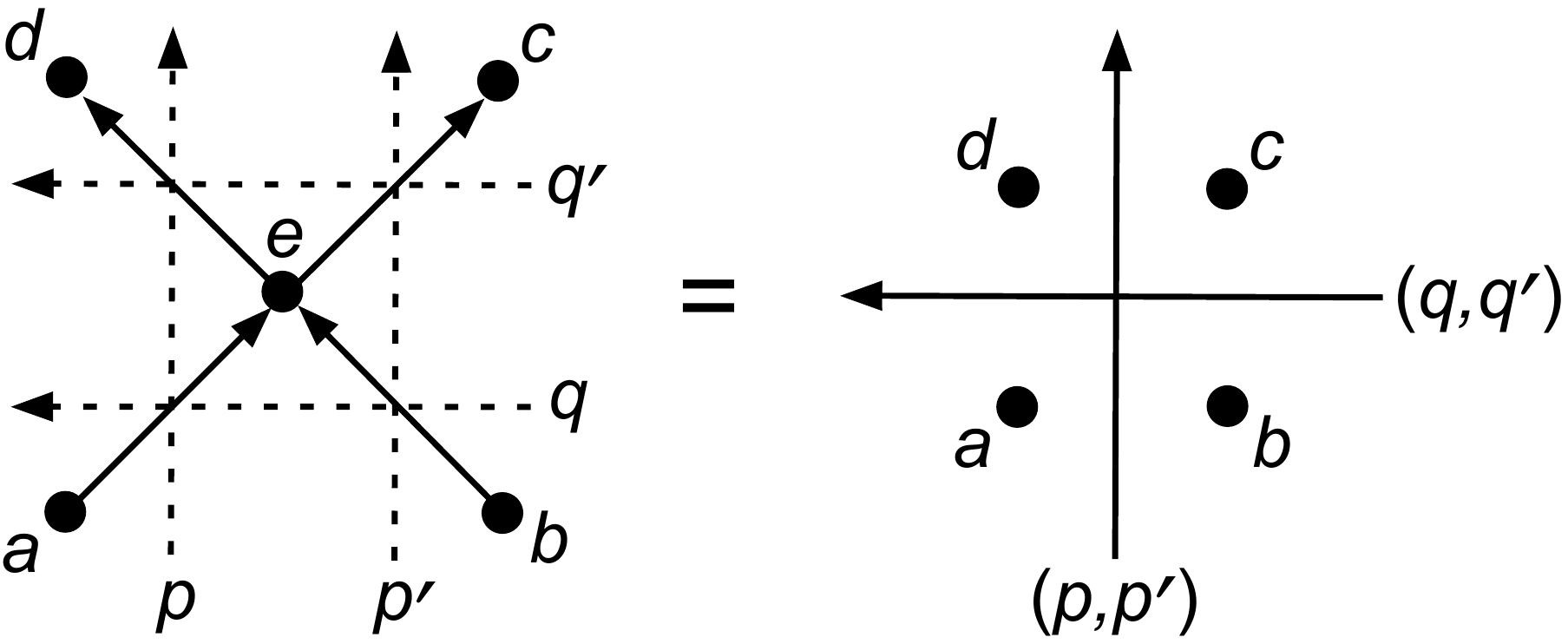}
\end{center}
\caption{Star of four chiral Potts weights (\ref{weights}) represented by
four oriented solid lines connecting the Potts spins $a$, $b$, $c$, $d$
and $e$. Summing over the value of $e$ this becomes the IRF model weight
on the right.}
\label{fig4}\end{figure}
 
In order to get a construction valid for all $N$, Baxter, Bazhanov
and Perk \cite{BBP} started in the opposite direction with a star
of chiral Potts model weights as in figure~\ref{fig4}, with Boltzmann
weights defined by (\ref{weights}) and figure~\ref{fig1}. Summing
out the central spin gives an Interaction-Round-a-Face (IRF) model
weight $U_{pp'qq'}(a,b,c,d)$, which can also be viewed as a vertex
model weight using the well-known map assigning spin differences
mod $N$ to the line pieces.\footnote{We could also have started with
a square as in figure~\ref{fig3}, as that setup differs by a Fourier
duality transform from the one in figure~\ref{fig4}.}

If one now chooses
$(x_{q'},y_{q'},\mu_{q'})=(y\vp_q,\omega^2 x\vp_q,\mu_q^{-1})$,
then $U_{pp'qq'}(a,b,c,d)=0$ for $0\le a-d\le1$ and $2\le b-c\le N-1$,
so that $U$ is triangular and the leading block is the R-matrix of
a $\tau_2$ model. If one also chooses
$(x_{p'},y_{p'},\mu_{p'})=(y\vp_p,\omega^2 x\vp_p,\mu_p^{-1})$,
then also $U_{pp'qq'}(a,b,c,d)=0$ for $0\le d-c\le 1$ and
$2\le a-b\le N-1$, and we receive a six-vertex R-matrix.

We consider from now on the superintegrable case
$(x_{p'},y_{p'},\mu_{p'})=(y\vp_p,x\vp_p,\mu_p^{-1})$. Then,
dropping some overall factors we can write the $2^2\!\times\!2^2$
R-matrix as
\begin{equation}
R(x,y)=\pmatrix{1-\XY\omega^{-1}&0&0&0\cr
0&1-\XY&\XY(1-\omega^{-1})&0\cr
0&1-\omega^{-1}&(1-\XY)\omega^{-1}&0\cr
0&0&0&1-\XY\omega^{-1}}
\label{6vR}\end{equation}
and the $2N\!\times\!2N$ $\tau_2$ R-matrix as
\begin{equation}
U(x)=\pmatrix{U_{00}&U_{01}\cr U_{10}&U_{11}}=
\pmatrix{{\bf1}-\omega x\bZ&-\omega x({\bf1}-\bZ)\bX\cr
\bX^{-1}({\bf1}-\bZ)&\omega(\bZ-x\mathbf{1})}.
\end{equation}
Here $x$ and $y$ are genus-0 rapidities of the six-vertex model,
$\bX$ and $\bZ$ are the $N\!\times\!N$ matrices defined before.

Note that $R(x,y)$ is a linear combination of only 1, $\xy$,
$\omega^{-1}$ and $\xy\omega^{-1}$. This makes that choice
particularly amenable for further analysis.

\subsection{The monodromy matrix and its expansion coefficients}

We can string $L$ $\tau_2$-model R-matrices together to form a
monodromy matrix.
Writing
\begin{equation}
{\cal L}(x)=\pmatrix{{\cal L}_{00}&{\cal L}_{01}\cr
{\cal L}_{10}&{\cal L}_{11}}=
\pmatrix{\A(x)&\B(x)\cr\C(x)&\D(x)},
\label{monod}\end{equation}
following notations of the Faddeev school \cite{STF}.
Then two such monodromy matrices ${\cal L}(x)$ and ${\cal L}(y)$
satisfy a Yang--Baxter equation with the six-vertex R-matrix.
This is so, both for the finite-dimensional case with $\omega$
and the infinite-dimensional case with generic $q$ that we shall
introduce later.

The monodromy matrix ${\cal L}(x)$ is a polynomial in $x$
of degree $L$. So, we write
\begin{equation}
\fl
\A(x)=\sum_{l=0}^L\mathbf{A}_lx^l,\quad
\B(x)=\sum_{l=1}^L\mathbf{B}_lx^l,\quad
\C(x)=\sum_{l=0}^{L-1}\mathbf{C}_lx^l,\quad
\D(x)=\sum_{l=0}^L\mathbf{D}_lx^l.
\end{equation}
In each of these four series all coefficients commute. It is
easy to work out some of these coefficients explicitly
\cite{AP2,AP5}
\begin{eqnarray}
\fl
\mathbf{A}_0=\mathbf{D}_L=\mathbf{1},\qquad
\mathbf{A}_L=\mathbf{D}_0\,\omega^{-L}=\prod_{j=1}^L\bZ_j,
\qquad\mathbf{C}_L=\mathbf{B}_0=0,
\nonumber\\ \fl
\overline\mathbf{B}_L={\mathbf{B}_L\over1-\omega}=
\sum_{j=1}^L\bigg(\prod_{m=1}^{j-1}\bZ_m\!\bigg)
{\hbox{\bfrak f}}_j,\qquad
\overline\mathbf{C}_0={\mathbf{C}_0\over1-\omega}=
\sum_{j=1}^L\bigg(\prod_{m=1}^{j-1}(\omega\bZ_m)\!\bigg)
{\hbox{\bfrak e}}_j,
\nonumber\\ \fl
\overline\mathbf{B}_1={\mathbf{B}_1\over1-\omega}=
\sum_{j=1}^L{{\bfrak f}}_j\prod_{m=j+1}^{L}(\omega\bZ_m),
\quad
\overline\mathbf{C}_{L-1}={\mathbf{C}_{L-1}\over1-\omega}=
\sum_{j=1}^L{{\bfrak e}}_j\prod_{m=j+1}^{L}\bZ_m.
\label{monoc}\end{eqnarray}
We note that some of these look like coproducts. The
${\bfrak e}_j$ and  ${\bfrak f}_j$ are defined by having
\begin{equation}
{\bfrak e}={\bX^{-1}({\bf1}-\bZ)\over1-\omega}\qquad\hbox{and}\qquad
{\bfrak f}={({\bf1}-\bZ)\bX\over1-\omega}
\end{equation}
acting only on position $j$.

The eigenstates of the superintegrable $\tau_2$ model are
highly degenerate and we can define creation and annihilation operators
within each sector by
\begin{equation}
\bx^+_{0,Q}=\mathbf{C}^{(N+Q)}_0\mathbf{B}^{(Q)}_1,\quad
\bx^-_{1,Q}=\mathbf{C}^{(Q)}_0\mathbf{B}^{(N+Q)}_1,\quad
(Q=0,\cdots ,N-1).
\end{equation}
Here we defined
\begin{equation}
\mathbf{B}^{(n)}_1={\mathbf{B}^{\;n}_1\over(n)_{\omega}!},\qquad
\mathbf{C}^{(n)}_0={\mathbf{C}^{\;n}_0\over(n)_{\omega}!},
\end{equation}
using the $q$-factorial and  $q$-integers
\begin{equation}
\fl
(n)_q!=(n)_q(n-1)_q\cdots(1)_q,
\qquad(n)_q={1-q^n\over1-q}=1+q+q^2+\cdots+q^{n-1},
\label{qint}\end{equation}
with $q=\omega$.

For each $Q$ between 0 and $N-1$, $\bx^+_{0,Q}$
and $\bx^-_{1,Q}$ generate an sl(2) (sub)algebra provided
the Serre relations
\begin{equation}
[\bx^+_{0,Q},[\bx^+_{0,Q},[\bx^+_{0,Q},\bx^-_{1,Q}]\,]\,]=0,\quad
[\bx^-_{1,Q},[\bx^-_{1,Q},[\bx^-_{1,Q},\bx^+_{0,Q}]\,]\,]=0,
\end{equation}
hold. 

\subsection{Infinite-dimensional representation}

In order to properly define all this, it will be important to
analytically continue in $\omega$. But then we need the infinite-dimensional
representation
\begin{equation}
\fl
\bX_q\equiv
 \pmatrix{     0&     0&       0&\ldots&           0&           0&\ldots\cr
               1&     0&       0&\ldots&           0&           0&\ldots\cr
               0&     1&       0&\ldots&           0&           0&\ldots\cr
          \vdots&\vdots&  \vdots&\ddots&      \vdots&      \vdots\cr
               0&     0&       0&\ldots&           0&           0&\ldots\cr
               0&     0&       0&\ldots&           1&           0&\ldots\cr
          \vdots&\vdots&  \vdots&      &      \vdots&      \vdots\cr},
\qquad\bZ_q\equiv
 \pmatrix{     1&     0&       0&\ldots&           0&           0&\ldots\cr
               0&     q&       0&\ldots&           0&           0&\ldots\cr
               0&     0&     q^2&\ldots&           0&           0&\ldots\cr
          \vdots&\vdots&  \vdots&\ddots&      \vdots&      \vdots\cr
               0&     0&       0&\ldots&\hskip-5pt
                              q^{N-2}\hskip-5pt&           0&\ldots\cr
               0&     0&       0&\ldots&           0&\hskip-20pt
                               q^{N-1}\hskip-20pt&\ldots\cr
          \vdots&\vdots&  \vdots&      &      \vdots&      \vdots\cr},
\end{equation}
satisfying $\bZ_q\bX_q=q\bX_q\bZ_q$.

Also, we really only need
\begin{equation}
{\bfrak e}_q={\bX_q^{-1}({\bf1}-\bZ_q)\over1-q}\qquad\hbox{and}\qquad
\hbox{\bfrak f}_q={({\bf1}-\bZ_q)\bX_q\over1-q}
\label{eqfq}\end{equation}
so that the extra 1, present in the finite-dimensional case
in the upper-right corner of $\bX$ and
making $\bX$ a cyclic matrix, cancels out.

Using the $q$-integers (\ref{qint}), we can write
\begin{equation}
\fl
{\bfrak e}_q\equiv
 \pmatrix{     0&\!\!\!(1)_q\!\!\!&  0&\ldots&     0&           0&\ldots\cr
               0&     0&\!\!\!(2)_q\!\!\!&\ldots&  0&           0&\ldots\cr
               0&     0&     0&\ldots&             0&           0&\ldots\cr
          \vdots&\vdots&  \vdots&\ddots&      \vdots&      \vdots\cr
               0&     0&       0&\ldots&     0& \!\!\!(n)_q\!\!\!&\ldots\cr
               0&     0&       0&\ldots&           0&           0&\ldots\cr
          \vdots&\vdots&  \vdots&      &      \vdots&      \vdots\cr},
\quad{\bfrak f}_q\equiv
 \pmatrix{     0&     0&       0&\ldots&           0&           0&\ldots\cr
     \!\!(1)_q\!\!\!& 0&       0&\ldots&           0&           0&\ldots\cr
          0&\!\!\!(2)_q\!\!\!& 0&\ldots&           0&           0&\ldots\cr
          \vdots&\vdots&  \vdots&\ddots&      \vdots&      \vdots\cr
               0&     0&       0&\ldots&           0&           0&\ldots\cr
               0&     0&       0&\ldots&  \!\!\!(n)_q\!\!\!&    0&\ldots\cr
          \vdots&\vdots&  \vdots&      &      \vdots&      \vdots\cr},
\label{efq}\end{equation}
or
\begin{equation}
({\bfrak e}_q)_{kl}=(k)_q\,\delta_{k+1,l},\qquad
({\bfrak f}_q)_{kl}=(l)_q\,\delta_{k,l+1},\qquad
(\bZ_q)_{kl}=q^{k-1}\,\delta_{k,l}.
\end{equation}

Also,
\begin{equation}
{\bfrak e}_q{\bfrak f}_q-q\,{\bfrak f}_q{\bfrak e}_q=
{1-q\bZ_q^{\;2}\over1-q},\qquad
{\bfrak e}_q\bZ_q=q\,\bZ_q{\bfrak e}_q,\qquad
\bZ_q{\bfrak f}_q=q\,{\bfrak f}_q\bZ_q.
\end{equation}

When $q$ is an $N$th root of unity, $(N)_q=0$ and the first $N\!\times\!N$
block decouples from the rest, making ${\bfrak e}_q$, ${\bfrak f}_q$
and $\bZ_q$ block diagonal. Said differently, replacing $q$ by $\omega$,
we only need to keep the first $N\!\times\!N$ block.

\subsection{Commutators and $q$-commutators}

Replacing $\omega$ by generic $q$ in (\ref{6vR}), the six-vertex
R-matrix with the special choice of the gauge rapidities in the
Yang--Baxter equation (see \cite[equation~(20)]{PA}) becomes
\begin{equation}
R(x,y)=\pmatrix{1-\XY q^{-1}&0&0&0\cr
0&1-\XY&\XY(1-q^{-1})&0\cr
0&1-q^{-1}&(1-\XY)q^{-1}&0\cr
0&0&0&1-\XY q^{-1}}.
\label{6vRq}\end{equation}
We can now take any monodromy matrix (\ref{monod}) associated with
it and write the Yang--Baxter equation out in terms of the
$\A$, $\B$, $\C$, $\D$ \cite{STF}. Then we get the following sixteen
equations:
\begin{equation}\fl
[\A(x),\A(y)]=[\B(x),\B(y)]=[\C(x),\C(y)]=[\D(x),\D(y)]=0,
\label{trivial}\end{equation}
\begin{eqnarray}
\fl(1-\xy\q)\B(x)\A(y)=\xy(1-\q)\B(y)\A(x)+(1-\xy)\q\A(y)\B(x),
\nonumber\\
\fl(1-\xy\q)\A(x)\B(y)=(1-\xy)\B(y)\A(x)+(1-\q)\A(y)\B(x),
\label{AB}\\
\fl\xy(1-\q)\C(x)\A(y)+(1-\xy)\A(x)\C(y)=(1-\xy\q)\C(y)\A(x),
\nonumber\\
\fl(1-\xy)\q\C(x)\A(y)+(1-\q)\A(x)\C(y)=(1-\xy\q)\A(y)\C(x),
\label{AC}\\
\fl\xy(1-\q)\D(x)\B(y)+(1-\xy)\B(x)\D(y)=(1-\xy\q)\D(y)\B(x),
\nonumber\\
\fl(1-\xy)\q\D(x)\B(y)+(1-\q)\B(x)\D(y)=(1-\xy\q)\B(y)\D(x),
\label{BD}\\
\fl(1-\xy\q)\D(x)\C(y)=\xy(1-\q)\D(y)\C(x)+(1-\xy)\q\C(y)\D(x),
\nonumber\\
\fl(1-\xy\q)\C(x)\D(y)=(1-\xy)\D(y)\C(x)+(1-\q)\C(y)\D(x),
\label{CD}\end{eqnarray}
\begin{eqnarray}
\fl\xy(1-\q)\D(x)\A(y)+(1-\xy)\B(x)\C(y)
\nonumber\\
=\xy(1-\q)\D(y)\A(x)+(1-\xy)\q\C(y)\B(x),
\nonumber\\
\fl\xy(1-\q)\C(x)\B(y)+(1-\xy)\A(x)\D(y)
\nonumber\\
=(1-\xy)\D(y)\A(x)+(1-\q)\C(y)\B(x),
\label{ABCD}\\
\fl(1-\xy)\q\D(x)\A(y)+(1-\q)\B(x)\C(y)
\nonumber\\
=\xy(1-\q)\B(y)\C(x)+(1-\xy)\q\A(y)\D(x),
\nonumber\\
\fl(1-\xy)\q\C(x)\B(y)+(1-\q)\A(x)\D(y)
\nonumber\\
=(1-\xy)\B(y)\C(x)+(1-\q)\A(y)\D(x).
\label{CDAB}\end{eqnarray}
Note that we have a symmetry under the simultaneous replacements
$\A\leftrightarrow\C$ and $\B\leftrightarrow\D$, as
(\ref{AB}) $\leftrightarrow$ (\ref{CD}) and
(\ref{AC}) $\leftrightarrow$ (\ref{BD});
if we replace $x\leftrightarrow y$ then also
(\ref{ABCD}) $\leftrightarrow$ (\ref{CDAB}).

At this point we stress that the special asymmetric gauge
of $R(x,y)$ causes all coefficients to be linear combinations
of only 1, $\xy$, $\omega^{-1}$ and $\xy\omega^{-1}$. This we
could have because we chose to use \cite{BBP} rather than \cite{BS}.
It makes expansions in powers of $x$ and $y$ particularly attractive.

Of course, the 16 equations (\ref{trivial}) through (\ref{CDAB})
with $x\leftrightarrow y$ are also valid. Using all 32 equations
we find the following 10 commutator equations
and 6 $q$-commutator equations:
\begin{eqnarray}
&[\A(x),\A(y)]=[\B(x),\B(y)]=[\C(x),\C(y)]=[\D(x),\D(y)]=0,
\nonumber\\
&[\A(x),\B(y)]=[\A(y),\B(x)],\quad[\C(x),\D(y)]=[\C(y),\D(x)],
\nonumber\\
&[\A(x),\D(y)]=[\A(y),\D(x)],
\nonumber\\
&y[\A(x),\C(y)]=x[\A(y),\C(x)],\quad y[\B(x),\D(y)]=x[\B(y),\D(x)],
\nonumber\\
&y[\A(x),\D(y)]+y[\B(x),\C(y)]=x[\A(y),\D(x)]+x[\B(y),\C(x)],
\nonumber\\
&[\C(x),\A(y)]_q=[\C(y),\A(x)]_q\,,\quad[\D(x),\B(y)]_q=[\D(y),\B(x)]_q\,,
\nonumber\\
&[\D(x),\A(y)]_q+[\C(x),\B(y)]_q=[\D(y),\A(x)]_q+[\C(y),\B(x)]_q\,,
\nonumber\\
&x[\A(x),\B(y)]_q=y[\A(y),\B(x)]_q\,,\quad x[\C(x),\D(y)]_q=y[\C(y),\D(x)]_q\,,
\nonumber\\
&x[\C(x),\B(y)]_q=y[\C(y),\B(x)]_q\,,
\end{eqnarray}
where $[\X,\Y]_q\equiv \X\Y-q\Y\X\;.$
This structure is a consequence of the special choice (\ref{6vRq}).

\subsection{Quantum group and coproducts}

The notion of quantum group came about after many years of progress
by many people. To me two papers by Zamolodchikov and Zamolodchikov
\cite{ZZ} and by Berg et al.\ \cite{BKWK} were very significant, as they
gave me the idea that somehow R-matrices are associated with groups.
At a conference in Kyoto, May 1981, Jimbo asked me how the models that 
I had presented in my talk \cite{PS} fit into a larger classification.
I answered him that I believed that they should be classified with
the first series of Dynkin diagrams.

Sklyanin seems to be the first to have suggested that the correct
mathematical framework associated with R-matrices is Hopf algebras,
in a one page note in Russian \cite{Sklyanin} on his two earlier
works on quantum algebra structures \cite{Sklyanin2,Sklyanin3}.
The works of Drinfeld \cite{Drinfeld1985} and Jimbo
\cite{Jimbo1985,Jimbo1986} describe a lot of the structure of
what now is called quantum groups, a term first coined by
Drinfeld \cite{Drinfeld1987}. Here I prefer to use Jimbo's review
\cite{Jimbo}, as it also addresses the chiral Potts model, albeit
only for $N$ odd, even though the case $N$ even can be dealt
with \cite{AP10} also.

To make contact with the quantum group
${\rm U}_q(\widehat{\hbox{\bfrak sl}}_2)$, we assume
$q\to\pm\omega^{1/2}$, defining a proper limiting
process where it is needed. We define the generators to be
\begin{eqnarray}
e_0=q\Z_q^{\,-1}\hbox{\bfrak f}_{q^2}\vp=
{\Z_q\vp-\Z_q^{\,-1}\over q-q^{-1}}\bX,\qquad
&e_1=f_0\lambda^{-1},
\nonumber\\
f_0=-q\hbox{\bfrak e}_{q^2}\vp\Z_q^{\,-1}=
-\bX^{\rm T}{\Z_q\vp-\Z_q^{\,-1}\over q-q^{-1}},\qquad
&f_1=e_0\lambda,
\nonumber\\
k_0=q\Z_q^{\,2},\qquad&k_1=k_0^{-1},
\end{eqnarray}
where $\lambda$ is a complex parameter and ${\bfrak e}_{q^2}$
and ${\bfrak f}_{q^2}$ are defined in (\ref{eqfq}). It is easily
verified that these generators satisfy the required relations
\begin{eqnarray}
e_if_j-f_je_i=\delta_{ij}{k_i\vp-k_i^{\,-1}\over q-q^{-1}},
\\
{e_i^{\,n}\over[n]_q!}\hbox{ and }\;{f_i^{\,n}\over[n]_q!}
\hbox{ Laurent polynomials in }q,
\label{Laurent}\\
k_i^{\,1/2}e_i\vp=qe_i\vp k_i^{\,1/2},\qquad
k_i^{\,1/2}e_j\vp=q^{-1}e_j\vp k_i^{\,1/2},
\nonumber\\
k_i^{\,1/2}f_i\vp=q^{-1}f_i\vp k_i^{\,1/2},\quad
k_i^{\,1/2}f_j\vp=qf_j\vp k_i^{\,1/2},
\\
e_i^{\,3}e_j\vp\,-[3]_q\,e_i^{\,2}e_j\vp e_i\vp\,
+[3]_q\,e_i\vp e_j\vp e_i^{\,2}\,-e_j\vp e_i^{\,3}\,=0,
\nonumber\\
f_i^{\,3}f_j\vp-[3]_q\,f_i^{\,2}f_j\vp f_i\vp
+[3]_q\,f_i\vp f_j\vp f_i^{\,2}-f_j\vp f_i^{\,3}=0,
\end{eqnarray}
where
\begin{equation}
[n]_q={q^n-q^{-n}\over q-q^{-1}}
\end{equation}
is the $q$-integer now. The condition (\ref{Laurent})
follows from (\ref{efq}) with
$(n)_q$ replaced by $[n]_q$.

Having two such representations we can prove that the following
coproduct satisfies the same relations:
\begin{eqnarray}
\Delta(e_0)=e_0\otimes q^{1/2}k_0^{1/2}+q^{-1/2}k_0^{-1/2}\otimes e_0,
\nonumber\\
\Delta(f_0)=f_0\otimes q^{-1/2}k_0^{1/2}+q^{1/2}k_0^{-1/2}\otimes f_0,
\nonumber\\
\Delta(e_1)=e_1\otimes q^{-1/2}k_1^{1/2}+q^{1/2}k_1^{-1/2}\otimes e_1,
\nonumber\\
\Delta(f_1)=f_1\otimes q^{1/2}k_1^{1/2}+q^{-1/2}k_1^{-1/2}\otimes f_1,
\nonumber\\
\Delta(k_0)=k_0\otimes k_0,\qquad\Delta(k_1)=k_1\otimes k_1.
\label{coprod}\end{eqnarray}
Thus this coproduct is indeed a quantum group homomorphism. By induction
we can define the coproduct $\Delta^{(L-1)}$ with $L$ factors
consistently. We only need to check the consistency for $\Delta^{(2)}$,
namely that we get equal results whether we replace the first factors
in (\ref{coprod}) by their coproduct or do this for the second factors.
It follows then that the coproduct $\Delta^{(L-1)}$ is also a quantum
group homomorphism.

Hence, realizing that the operators in (\ref{monoc}) are coproducts,
we can use this fact to greatly simplify checking their relations,
as it is sufficient to check them for $L=2$ or 3 only. This can be used
also for proving higher Serre relations in \cite{AP5}, for example,
to which we may return in a future paper.

\ack
It is a great pleasure to thank Professor Rodney Baxter for many years
of warm friendship and encouragement. His papers and the many discussions
together with him, going as far back as 1976, were highly appreciated
and formative. This work would not have been possible without the many
contributions of Dr. Helen Au-Yang Perk. Her imprint is on about every
page of this paper.
The author also thanks Professors Vladimir Bazhanov, Murray Batchelor,
and Vladimir Mangazeev for their kind invitation and warm hospitality
and for providing the financial support for him to attend the special
meeting at Palm Cove, where an early account of this paper was
presented.


\end{document}